\documentclass[letter]{aa}  

\usepackage{graphicx}
\usepackage{txfonts}
\usepackage[colorlinks=true,citecolor=blue]{hyperref}
\usepackage{natbib}
\newcommand{\micron}{\,\mu\mathrm{m}}
\newcommand{\microJybm}{\,\mathrm{\mu Jy\,beam^{-1}}}
\newcommand{\microJy}{\,\mathrm{\mu Jy}}
\newcommand{\ergs}{\,\mathrm{erg\,s^{-1}}}

\newcommand{\WHz}{\,\mathrm{W\,Hz^{-1}}}

\begin{document}

   \title{The radio emission from active galactic nuclei}

   \author{J. F. Radcliffe\inst{1,2,3}
           \and P. D. Barthel\inst{1} 
           \and M. A. Garrett\inst{3,4}
           \and R. J. Beswick\inst{3}
           \and A. P. Thomson\inst{3}
           \and T. W. B. Muxlow\inst{3}}

   \institute{Kapteyn Astronomical Institute, University of Groningen, 9747 AD Groningen, The Netherlands \\
        \email{pdb@astro.rug.nl}
        \and Department of Physics, University of Pretoria, Lynnwood Road, Hatfield, Pretoria, 0083, South Africa
        \and Jodrell Bank Centre for Astrophysics, School of Physics \& Astronomy, The University of Manchester, Alan Turing Building, Oxford Road, Manchester M13 9PL, UK
        \and Leiden Observatory, Leiden University, PO Box 9513, 2300 RA Leiden, The Netherlands}

   \date{Received xxx; accepted xxx}

% \abstract{}{}{}{}{} 
% 5 {} token are mandatory
 
  \abstract
   {For nearly seven decades, astronomers have been studying active galaxies, that is to say, galaxies with actively accreting central supermassive black holes: active galactic nuclei (AGN). A small fraction are characterized by luminous, powerful radio emission: This class is known as radio-loud AGN. A substantial fraction, the so-called radio-quiet AGN population, display intermediate or weak radio emission. However, an appreciable fraction of strong X-ray-emitting AGN are characterized by the absence of radio emission, down to an upper limit of about $10^{-7}$ times the luminosity of the most powerful radio-loud AGN.}
   {We wish to address the nature of these -- seemingly radio-silent -- X-ray-luminous AGN and their host galaxies to determine if there is any radio emission, and, if so, where it originates.}
   {Focusing on the GOODS-N field, we examine the nature of these objects, employing stacking techniques on ultra-deep radio data obtained with the JVLA. We combine these radio data with \textit{Spitzer} far-infrared data.}
   {We establish the absence, or totally insignificant contribution, of jet-driven radio emission in roughly half of the otherwise normal population of X-ray-luminous AGN, which appear to reside in normal star-forming galaxies.}
   {AGN- or jet-driven radio emission is simply a mechanism that may be at work or may be dormant in galaxies with actively accreting black holes. The latter cases can be classified as radio-silent AGN.}


   \maketitle
%
%-------------------------------------------------------------------

\section{Introduction }

Emitting hundreds to thousands of radio flux units (janskys), active galaxies and active galactic nuclei (AGN) were originally discovered through their luminous radio emission. It was, however, realized early on that radio loudness in active galaxies is the exception rather than the rule. For instance, many optically identified quasi-stellar objects (QSOs) have been reported \citep{sandage1965} to show no sign of radio emission. Following decades of research into the nature of this radio-quiet QSO population, a recent ultra-deep study by \citet{Kellermann2016:RQAGN} suggested that the 6 GHz radio luminosity function of optically selected low-redshift Sloan Digital Sky Survey (SDSS) QSOs primarily comprises two components. The first is the AGN, jet-driven component that smoothly covers the (6\,GHz) radio luminosity range from about $10^{27}$ down to $10^{23}\WHz$. The second, which is prominently present in the sample, is the QSO host galaxy, starburst-driven radio emission that covers the $10^{21}\mbox{--}10^{23}\WHz$ range. The latter emission is indicative of the formation of a few to several tens of solar-mass stars per year. The host galaxy is responsible for the dominant radio emission component in approximately 80\% of the sample: These objects make up the so-called radio-quiet QSO population. Lacking sufficient angular resolution, the standard connected interferometers, such as the Very Large Array (VLA) and the now-operational Square Kilometre Array (SKA) precursors, are unable to quantify any low-level AGN-driven radio emission in these radio-quiet objects, that is to say, jet-driven radio emission in the range $10^{21}\mbox{--}10^{22}\WHz$. For nearby low-luminosity AGN, that decomposition can successfully be made by exploiting the radio--far-infrared (FIR) correlation \citep[e.g.,][]{wilson1988,barthel2006}, but the multi-faceted physical origin of the radio emission in the radio-quiet AGN population is still a matter of debate \citep{panessa2019}. Here we examine these extragalactic radio populations in more detail.  

The extragalactic radio sky is currently studied down to microjansky depths \citep[e.g.,][]{smolcic2017,owen2018,mauch2020}. The sky surface density of the ultra-luminous Third Cambridge Catalogue of Radio Sources (3C) objects is about one per 10 $\times$ 10 degrees. These radio sources, having projected dimensions up to hundreds of kiloparsecs, reach radio luminosities of up to $10^{28}\WHz$ (at $1.4\,\mathrm{GHz}$) at redshifts, $z$, of up to 2.8. They are all identified with powerful AGN: radio galaxies and quasars. Another well-known radio survey, the NRAO VLA Sky Survey \citep[NVSS;][]{condon1998}, goes about a factor of a thousand deeper, to the millijansky regime, and shows a source surface density of about one per $10^\prime\times 10^\prime$, with sources with radio luminosities down to about $10^{23}\WHz$. It is thought that the large majority of these radio sources are AGN-driven. That situation changes when we go down another factor of a hundred in depth. At arcsecond angular resolutions, the radio source surface density at the $10\microJy$ level is 2--4 per square arcminute. Slightly resolved starburst-driven radio sources, displaying alignment with their host galaxies and producing $10^{21}\mbox{--}10^{23}\WHz$, outnumber the AGN-driven objects at that level \citep[e.g.,][]{muxlow2005high,ibar2008,padovani2009,bonzini2013,barger2015,barger2017,owen2018,ceraj2018,mauch2020}. This starburst dominance is also clearly seen as a submillijansky upturn in the radio source counts \citep{smolcic2017}, as well as in the ultra-deep radio survey of the general QSO population, as previously discussed. Hence, identifying AGN in the submillijansky radio population is not straightforward: They may occur at a low level in hybrid systems or be absent.

Earlier, our team addressed these issues using the unique capacities of very long baseline interferometry (VLBI). \citet{Radcliffe2018:p1} carried out ultra-deep wide-field VLBI in the GOODS-N (HDF-N) field and reported the identification of 31 AGN in the redshift range 0.11--3.44 among hundreds of radio sources in that well-studied field. The Radcliffe et al. (in press) follow-up study describes the nature of the host galaxies of the AGN and hybrid AGN--starburst systems: (1) AGN-dominated hybrid systems with efficient accretion, (2) intermediate redshift early-type hosts with inefficient AGN accretion, and (3) very dusty hybrid systems at high redshift, such as the well-known submillimeter source GN16. 

Another important result of the Radcliffe et al. (in press) study is that about one-third of the VLBI AGN remain undetected in the deep $\mathrm{2\,Ms~ 0.5\mbox{--}7\,keV}$ \emph{Chandra} X-ray observations \citep{xue_xray_2016}. That fraction is in rough agreement with the fractional occurrence -- or, more accurately, the absence -- of X-ray emission in infrared-selected QSOs, as reported by \citet{DelMoro2016} and \citet{Mateos:2017fm}. Confirming the \citet{Padovani2016} results, these X-ray-deficient AGN occur in both distant passive, inefficiently accreting systems and in very dusty host galaxies, which renders them weak or Compton-thick. The recent multiwavelength study by \citet{lambrides2020} underlined the incompleteness of AGN X-ray studies. 

However, while incomplete for certain AGN, X-ray surveys still yield the highest AGN\footnote{Defined primarily via their wideband X-ray luminosity (at known redshift): $L_X > 3\times10^{42}\,\mathrm{erg\,s^{-1}}$, or $> 3\times10^{35}\,\mathrm{W}$. This is not a robust lower limit since (rare) luminous distant starburst galaxies are known to produce $\sim 10^{43}\,\mathrm{erg\,s^{-1}}$ \citep{barger2017}. This surface density should be compared to that of faint galaxies, which is $\sim 1700$ per square arcmin, as inferred for the Hubble Ultra Deep Field \citep{beckwith2006}.} surface density, of about 7 per square arcmin \citep{Luo2017:xray}. As reviewed by \citet{Brandt2015}, X-ray surveys have been of the utmost importance in AGN research, but the enigmatic aspect of the large X-ray surface density has not yet been resolved: It remains unclear what the nature of the luminous X-ray AGN without detectable radio emission is. The issue is nicely illustrated by the detection statistics. \citet{barger2017} find that the faint radio and deep \emph{Chandra} X-ray populations are two disjunct populations, with only a certain level of overlap. Of the 445 1.4\,GHz radio sources in the central 124 square arcmin of GOODS-N (AGN, starbursts, and mixed systems), 31\% have X-ray counterparts (within 1\farcs5) and 69\% have no X-ray counterparts. Of the GOODS-N X-ray sources down to $f_{0.5-2{\rm keV}} \approx 1.5\times10^{-17}\,\mathrm{erg\,cm^{-2}\,s^{-1}}$, only 51\% are associated with a ($>11.5\microJy$) radio source. We focus here on the other half of the X-ray AGN -- those that remain undetected in the radio.

\section{Methods and results: Radio-silent AGN}

To investigate the nature of the ultra-weak (or absent) radio emission of these X-ray-bright sources, we carried out stacking techniques in our database for GOODS-N. We took advantage of the \citet{owen2018} ultra-deep ($1.8 \microJybm$) 1--2 GHz VLA observations, obtained as part of the \emph{e}-MERGE survey, that have a resolution of approximately 1\farcs6 \citep{muxlow2020}. There are 334 non-stellar sources in the 2\,Ms \emph{Chandra} GOODS-N catalog with positional accuracy better than 0\farcs5 \citep{xue_xray_2016} that are also located within the \emph{Spitzer}-Multiband Imaging Photometer (MIPS) $24\micron$ field of view. Of these, 168 sources (50.3\%) have no VLA radio counterparts above a mean five-sigma detection threshold of $10\microJybm$, within a 1\farcs5 search radius. This detection fraction is entirely consistent with the results of \citet{barger2017}. To isolate a radio-silent population, we further filtered these 168 non-detections using a median absolute deviation (MAD) outlier filter \citep{iglewicz1993detect}. For each pixel, $x_i$, in each $80\times 80$ pixel radio image, we computed the modified $Z$-score, $M_i$, using
\begin{equation}
         M_i  = 0.6745 \frac{(x_i-\tilde{x})}{\rm MAD},
\end{equation}
 where $\mathrm{MAD} = \mathrm{median}(\left|x_i-\tilde{x}\right|)$ and $\tilde{x}$ is the median of all the pixels in the image. Pixels were defined as outliers if $\left|M_i \right| > 3$, and a source image was rejected if there was an outlier pixel within 1.5 times the size of the VLA restoring beam ($\sim 2^{\prime\prime}$) of the X-ray position. This technique removed 78 sources with probable VLA counterparts with S/N of 3-5, that is to say, peak brightness values of 6--10$\microJybm$. Finally, we excluded two sources that are behind extended radio emission from an FR-I radio galaxy. This yields a final sample of 88 X-ray sources (with known redshifts), or 26.3\% of the original sample, that have a low probability of a radio counterpart. The redshift range of this sample is 0.079--5.186, with a mean value of 1.54.

\begin{figure*}
        \centering
        \includegraphics[width=\linewidth]{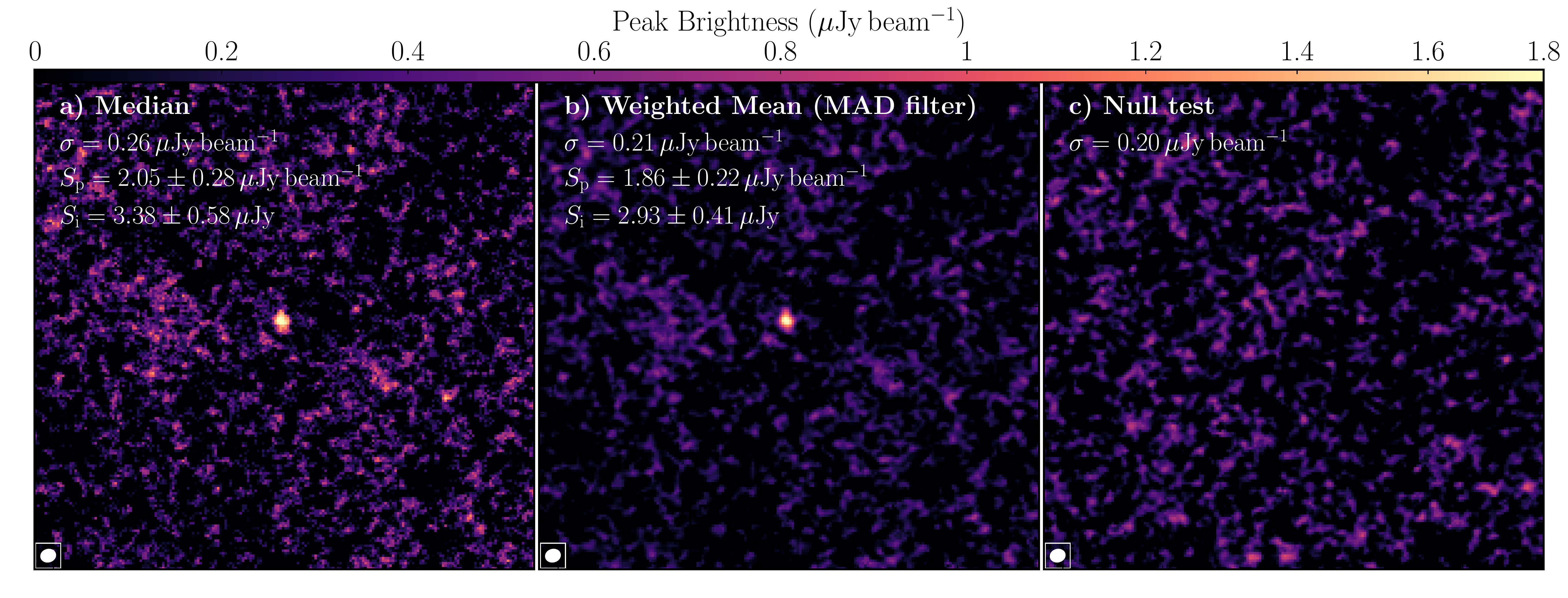}
        \caption{Stacking results for 88 objects with X-ray luminosities between $1.9\times 10^{39}$ and $4.1\times10^{44}\ergs$. The images measure $28^{\prime\prime} \times 28^{\prime\prime}$, with an angular resolution of $\sim 2^{\prime\prime}$.}
        \label{fig:stack_results}
\end{figure*}

For each of these 88 X-ray sources, an $80 \times 80$ pixel ($28^{\prime\prime} \times 28^{\prime\prime}$) cutout centred on the X-ray position was excised. These images were aligned and a median and a weighted mean stack (Fig.~\ref{fig:stack_results}a,b) were performed on a pixel-by-pixel basis. For the weighted mean stack, we added another MAD filter to identify and remove nearby sources. The weights used per image were proportional to the inverse square of the local rms noise. To ensure that the stacking routine was robust, we performed a null test. For the same number of stacks, we added an additional random term (between $-15^{\prime\prime}$ and $+15^{\prime\prime}$) to the X-ray positions, and, as Fig. \ref{fig:stack_results}c shows, the stacking signal ceased to exist.

The median and weighted mean stacks in Fig.~\ref{fig:stack_results} clearly show significant radio emission, with a S/N greater than 10. They correspond to just over two times the noise in the individual images, and the flux density of the median stack is approximately 12\% higher than that of the mean stack. While these are within the error bounds, it implies that the underlying distribution is slightly skewed: There must be more stronger radio sources than weaker ones.

We therefore performed additional stacks, binning the data into two X-ray luminosity groups (which are essentially also redshift groups). As shown in Table~\ref{tab:stack_results}, the radio flux density changes only mildly with X-ray luminosity and corresponds to radio powers around $10^{21}\mbox{--}10^{22}\WHz$. This is in the star-formation-dominated luminosity regime \citep[e.g.,][]{smolcic2017}. To test whether the radio emission in the stacks indeed originates from star formation, we used the \emph{Spitzer} MIPS $24\micron$ flux to compute $q_{24}$ parameter values in the FIR-radio diagnostic \citep[e.g.,][]{Appleton24um2004,ibar2008}. Nearly all sources (84\%) have $24\micron$ counterparts with flux densities between 20 and 80$\microJy$, with around 10\% of sources in the range 100--300 $\microJy$. These excess sources affect the mean flux density of this distribution, so we instead used the median $24\micron$ values. In order to establish these values for each X-ray luminosity bin, we had to take the non-detections into account. We employed left-censored survival analysis, using a Kaplan-Meier estimator to estimate the median and 95\% confidence intervals of the distribution. The results, including the inferred values of radio luminosity and star-formation rates (SFRs), are shown in Table~\ref{tab:stack_results}. 

\begin{table*}
        \centering
        \caption{Stacking results for two $L_X$ groups of the 88 radio-undetected objects. The 1.5\,GHz stack flux densities represent peak values, and the radio luminosity computation assumed a radio spectral index of $-0.7$. The SFR computation, following \citet{Novak2017:co}, used these luminosity values. The resulting SFR values represent upper limits (see the main text).}
        \label{tab:stack_results}
        \begin{tabular}{ccccccc}
        \hline\hline
        $\mathrm{L}_{0.5-7\,\mathrm{keV}}$ & $N_s$ & $z_\mathrm{median}$ & $S_\mathrm{1.5\,GHz}$ & $q_{24}$ & $L_\mathrm{1.5\,GHz}$ & SFR \\
         ($\mathrm{erg\,s^{-1}}$) & & & ($\mathrm{\mu Jy\,beam^{-1}}$) & & ($\mathrm{W\,Hz^{-1}}$) & ($\mathrm{M_\odot\,yr^{-1}}$) \\
        \hline 
        $1.9 \times 10^{39}-1.0 \times 10^{43}$ & 43 & 0.75  & $1.4\pm0.36$ & $1.10\substack{+0.27\\-0.34}$ & $3.2\times 10^{21}$ & $\lesssim 1.2$ \\
        $1.0 \times 10^{43}-4.1 \times 10^{44}$ & 45 & 2.18 & $2.5\pm0.45$ & $0.89\substack{+0.43\\-0.20}$ & $6.6\times 10^{22}$ & $\lesssim 15$ \\
        \hline
        \end{tabular}
\end{table*}

As judged from Table~\ref{tab:stack_results}, the $q_{24}$ parameter is in the middle of the star-formation-dominated regime ($0.2 < q_{24} < 1.5$), both at low and high X-ray luminosities; within the errors, they are the same. We tested the sensitivity to redshift of the median $q_{24}$ values for stacks in different redshift intervals and found no effect. We observe no sign of any AGN-driven radio excess. Recalling the cosmologically remarkable stable $q_{24}$ parameter as measured for star-forming galaxies over a wide range of redshift by \citet{Appleton24um2004}, we therefore suggest that star formation, with inferred rates of a few to some tens of solar masses ($\rm M_\odot$) per year, is responsible for (the bulk of) the radio emission of the (sub)microjansky population of X-ray-selected AGN. These objects are ``non-jetted'' AGN in normal star-forming galaxies. We caution nevertheless that we cannot exclude a small level of AGN emission in the radio or mid-IR bands: The SFR estimates in the last column of Table~\ref{tab:stack_results} must be considered as upper limits.

Hereafter, defining the non-jetted AGN class as having $L_\mathrm{jet}$ of at most 10\% of $L_\mathrm{stack}$ ($L_\mathrm{1.5\,GHz}$), which translates into $< 3\times10^{20}\WHz$ at $z=0.8$ or $ < 7\times 10^{21}\WHz$ at $z=2.2$, we infer that any AGN-driven (or better, jet-driven) radio emission must amount to less than about a few times $10^{21}\WHz$. This represents a factor of about seven orders of magnitude in comparison with the most luminous radio-loud AGN. Given that the $q_{24}$ values are seen not to change with X-ray luminosity (redshift), whereas the radio luminosities do, we measure higher SFRs at higher redshift in these AGN hosts -- a well-known fact for non-active galaxies \citep[e.g.,][]{whitaker2012} and for distant radio-loud 3CR objects \citep{Podigachoski2015}.

\begin{figure}
        \centering
        \includegraphics[width=\linewidth]{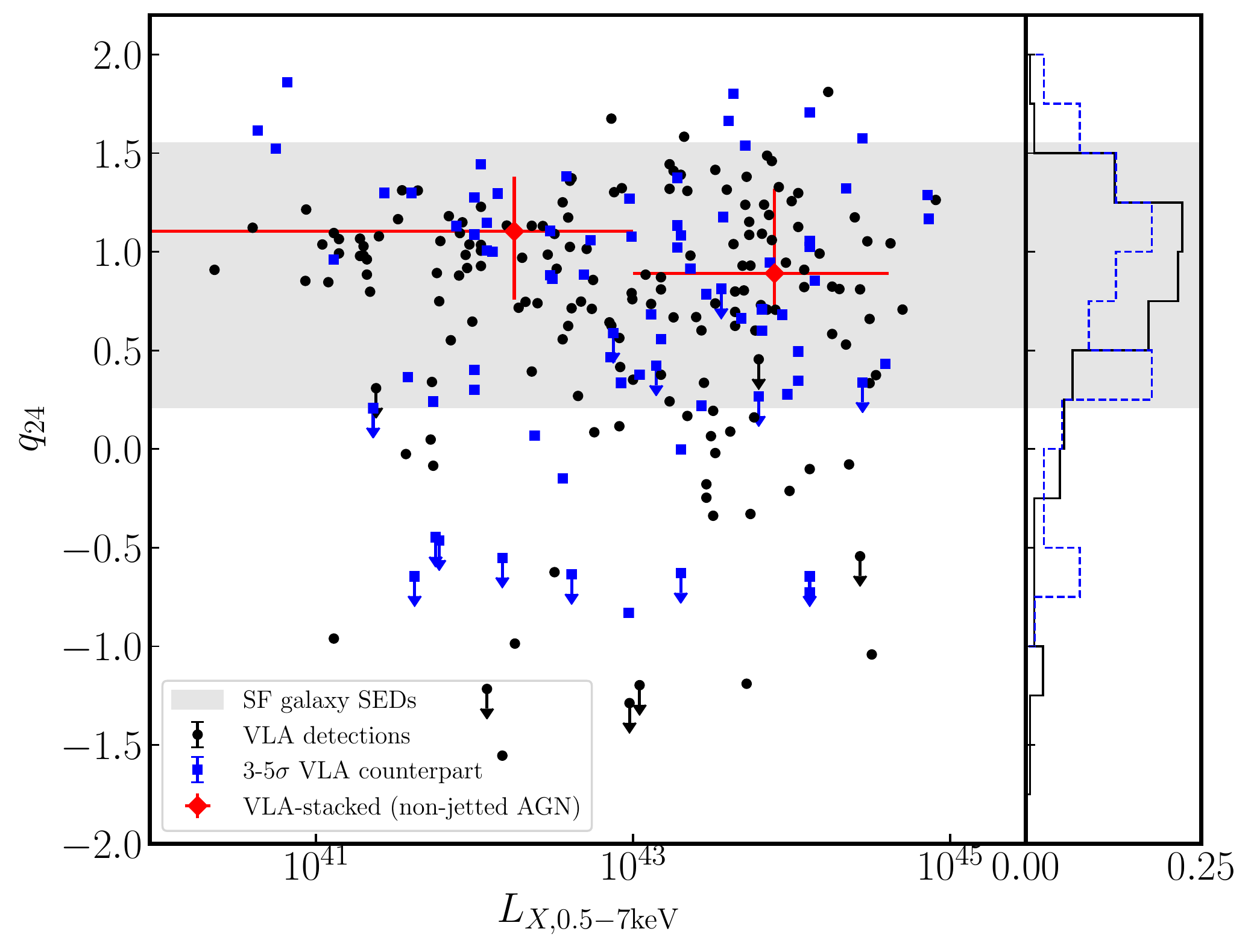}
        \caption{The FIR/radio ratio, $q_{24}$, as a function of the X-ray luminosity for three groups of X-ray AGN: $>5$ sigma VLA detections (black), $3\mbox{--}5$ sigma VLA detections (blue), and radio-undetected (red); the normalized frequency distributions appear at the right. The gray bar represents the range of $q_{24}$ between redshifts of $0 \leq z \leq 3$ for five star-forming galaxy templates \citep[data acquired from][]{delmoro2013}.}
        \label{fig:radio_excess}
\end{figure}

Figure~\ref{fig:radio_excess} shows the distribution of the $q_{24}$ values for the 78 individually detected radio sources in the 3--5 sigma set and the 166 objects in the $>5$ sigma set together with the two stack values for the radio-undetected X-ray AGN. The gray band indicates the star-formation-dominated regime; the area under it is commonly known as the radio-excess region. As seen from Table~\ref{tab:stack_results}, the mean $q_{24}$ value for the stacked objects is 0.99. The distributions of the $q_{24}$ values for the $>5$ sigma and the 3\mbox{--}5 sigma sets are statistically indistinguishable, but the $q_{24}$ values are significantly lower: The combined mean value is 0.73. It is clear that the range of $10\microJy\mbox{--}1\microJy$ marks the transition of jet-dominated to star-formation-dominated X-ray sources: At this point, any radio emission is completely dominated by star-formation-related radio processes. Additionally, the radio emission of substantial fractions of the $>5$ sigma and $3\mbox{--}5$ sigma sets must be largely star-formation-driven. It is noteworthy that this microjansky transition does not come as a surprise since the associated SFRs are entirely as expected \citep{whitaker2012}. This result recalls earlier work by \citet{Richards2007}: From considerably shallower radio data, they conclude that a large fraction of X-ray AGN appear as starbursts when using radio diagnostics. 
The $L_\mathrm{jet}$ upper limits mentioned above are in the range of the faint unresolved radio cores observed \citep{ho2008} in low-luminosity AGN\footnote{But still orders of magnitude more luminous than the ultra-compact radio nucleus in the Galactic Center, Sgr A* \citep{genzel2010}.} in the local Universe. We conclude that at least one-quarter (the stacked set) but probably more like half of the extragalactic X-ray population has no, or totally insignificant, relativistic radio jets with associated AGN-driven radio emission. Following an earlier proposal \citep{Padovani2017}, we classify the radio-silent AGN as non-jetted AGN. This type of object has an accretion disk and a hot corona, and hence luminous X-ray emission, but simply does not form radio jets of any significance.

\section{Discussion: The AGN radio source mechanism}

\begin{figure}
        \centering
        \includegraphics[width=\linewidth]{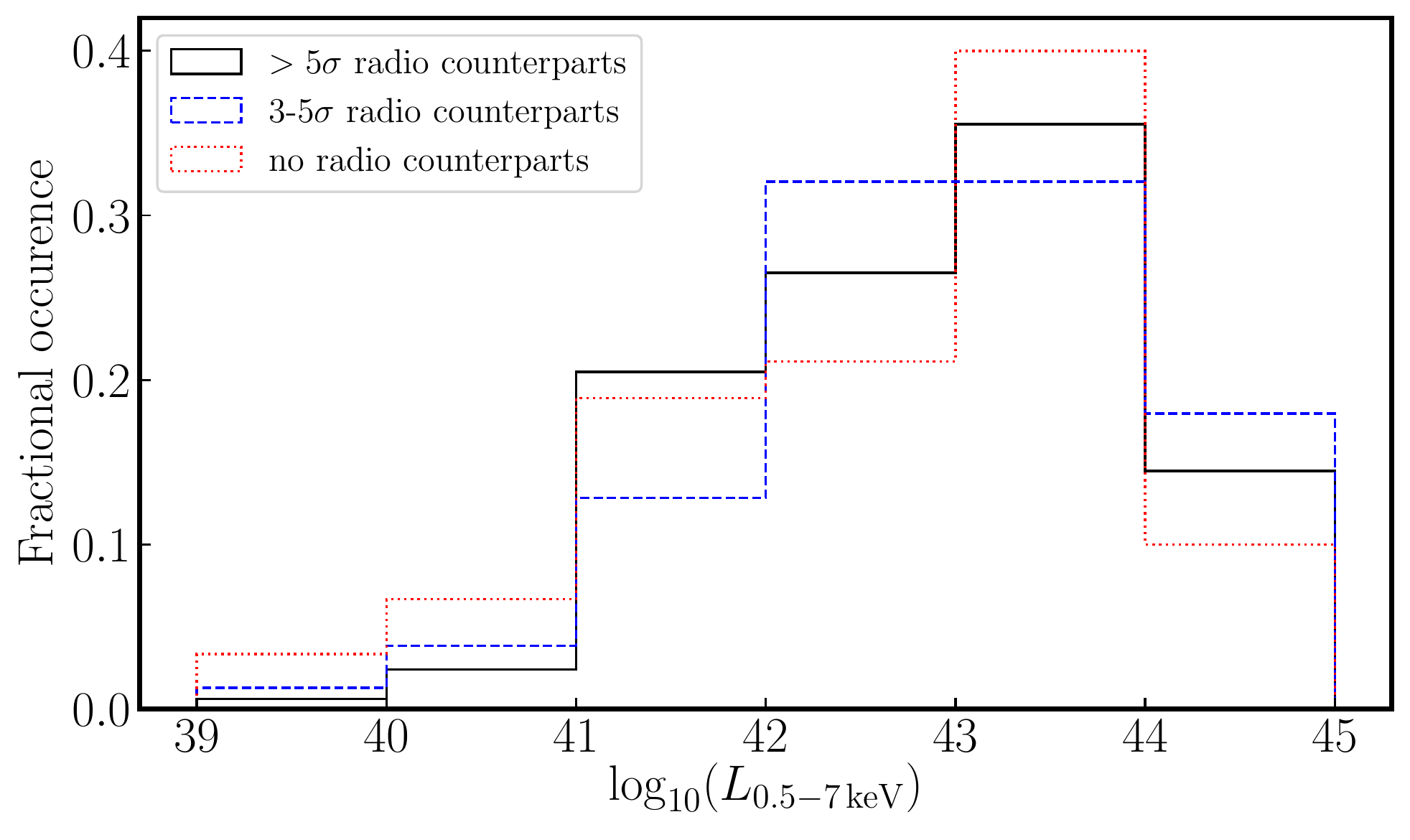}
        \caption{Histograms of $L_X$ relative frequency for three radio groups: jet-dominated (black), mixed systems (blue), and non-jetted (red).}
        \label{fig:luminosity_histo}
\end{figure}

We proceeded by examining the occurrence of non-jetted AGN in relation to their AGN strength. In Fig.~\ref{fig:luminosity_histo} we present histograms of the luminosities of three groups of X-ray-selected objects: (1) 166 objects in excess of 10 $\microJy$ (which are the jet-dominated but still hybrid starburst-AGN group); (2) 78 objects that have radio emission in the $5\mbox{--}10\microJy$ range (i.e., the hybrid starburst-AGN group); and (3) the 88 objects without VLA radio emission (i.e., the [stack] radio-silent group).

\begin{figure}
        \centering
        \includegraphics[width=\linewidth]{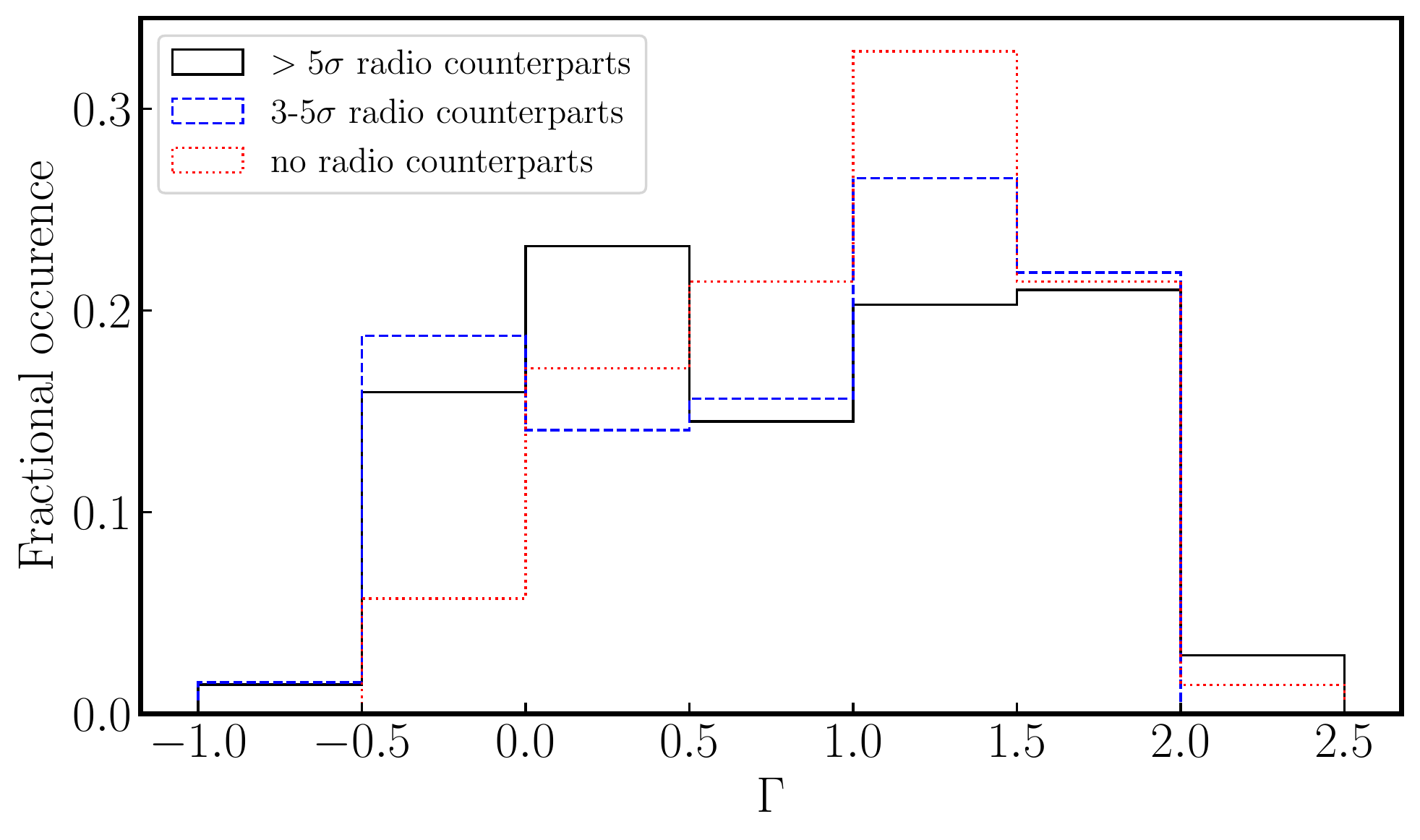}
        \caption{Histograms of $\Gamma$ relative frequency for three radio groups: jet-dominated (black), mixed systems (blue), and non-jetted (red).}
        \label{fig:gamma_hist}
\end{figure}

Kolmogorov-Smirnov (K-S) tests indicate that these distributions, which cover six orders of magnitude in X-ray luminosity, are drawn from the same population, at 99\% confidence. This also reflects the fact that the redshift distributions of the three samples are virtually identical. We therefore conclude that X-ray AGN may, or may not, develop jets producing significant radio emission, the strength of which is unrelated to the X-ray strength\footnote{We omit the rare radio-loud blazar class, in which Doppler-boosted jet emission is responsible for strong, variable, correlated emission at all wavelengths, from the present discussion.}. AGN-driven radio emission -- jets, on parsec to hundreds of kiloparsec scales, plus the large-scale radio lobes that they feed -- must be simply a special mechanism that is unrelated to the strength of the central supermassive black hole accretion as manifested in its emission at other wavelengths. We stress, however, that some degree of correlation between AGN jet strength and X-ray strength, as a function of black hole accretion efficiency, is likely. This is illustrated, for instance, by the fact that the X-ray luminosities of the ultra-powerful, efficiently accreting 3C sources reach $10^{46}\mathrm{erg\,s^{-1}}$ \citep{wilkes2013}, while their histogram shape is similar to those in Fig.~\ref{fig:luminosity_histo}. We furthermore conclude that the level of host starburst contamination at low X-ray luminosities must be similar for the three groups (and we recall the fact that star-forming galaxies have been reported by \citealt{barger2017} to display correlated radio and X-ray luminosities, up to $L_X \approx 10^{42}\ergs$). X-ray-emitting AGN appear to be hardly affected by the presence or the strength of jet-driven radio emission: The two mechanisms are unrelated. Most interestingly, the distributions of the X-ray spectral slopes for the three groups are not exactly the same. Figure~\ref{fig:gamma_hist} shows the effective X-ray photon indices. The radio-silent group peaks at $\Gamma=1.4$ and shows only a modest tail toward harder photon indices. This is in contrast to the radio-quiet group and particularly the radio-loud group, as confirmed with K-S tests. They show pronounced tails of hard photon indices. Such hard indices are generally attributed to the effect of radio jets \citep{wilkes1987,zhu2020}, or they could indicate different host interstellar medium (absorption) properties \citep{Brandt2015}, which will clearly be an issue for further study.

Our findings are obviously relevant for the local AGN picture, where low-luminosity AGN often do not have relativistic radio jets \citep{ho2008,Padovani2017}, and, as mentioned earlier, they provide strong support for the jetted versus non-jetted scenario \citep{Bonzini2015,Padovani2017}. Our presented results have interesting implications. Firstly, the positive stacking signal suggests that many X-ray AGN will have radio counterparts in the nanojansky sensitivity surveys that will be provided by the SKA and next-generation Very Large Array. Secondly, deeper, more sensitive VLBI observations would be needed to discover radio jets in those objects at the level below $10^{21}\WHz$. Future surveys will require both deep X-ray and radio observations (subarcsecond as well as milliarcsecond resolution) in order to separate the jetted from the non-jetted AGN and to characterize the hybrid AGN-starburst systems. Following up on \citet{Garofalo2019} and \citet{zhu2020}, a multispectral approach will be crucial to understanding the conditions under which jets will, or will not, develop. Finally, ultra-deep optical and infrared observations will be needed to investigate the possible occurrence of non-jetted Compton-thick AGN.

\section{Conclusions}

At the microjansky level, radio surveys run into many star-forming galaxies with SFRs of a few to a few tens of solar masses per year; AGN-driven jets contribute negligible amounts of radio emission in these objects. In cases where these jet contributions are non-negligible, a radio excess is apparent, classifying these objects as radio-loud AGN. However, X-ray AGN without radio jets also exist over a wide range of X-ray luminosities. They represent at least one-quarter and probably half of the extragalactic X-ray population. We have identified their nature: They are coreless and jet-less (or non-jetted) accretors in massive star-forming galaxies, in which all or virtually all radio emission draws from star-formation-related processes in the AGN host galaxy. Whereas their host galaxies generate radio emission, their AGN contribute negligibly -- they are radio-silent. 

\begin{acknowledgements}
The GOODS-N radio sky was addressed in the 2019 PhD Thesis work of author Radcliffe, carried out with the team of co-authors. He and co-author Barthel subsequently examined the radio-invisible population in that field. Barthel wrote the first version of the Article, to which Radcliffe added discussion and diagrams, and to which the other co-authors contributed in its final stage. Barthel and Radcliffe acknowledge useful discussions with Karina Caputi, Luis Ho, Joanna Kuraszkiewicz, Paolo Padovani, and Belinda Wilkes. A useful referee report is also acknowledged.
\\
This research made use of Astropy, a community-developed core Python package for Astronomy \citep{Astropy,Astropy2018}. It has received funding from the European Commission Seventh Framework Programme (FP/2007-2013) under grant agreement No 283393 (RadioNet3). J.F.R. acknowledges the Science and Technologies Facilities Council (STFC), an Ubbo Emmius scholarship from the University of Groningen, and the South African Radio Astronomy Observatory (SARAO) whose funding contributed to this research.
\end{acknowledgements}
\bibliographystyle{aa} % style aa.bst
\bibliography{Radio_silent_AGN} % your references Yourfile.bib

\end{document}